\documentclass[manuscript]{aastex}     
\usepackage{url}
\shorttitle{SUB-SURFACE MERIDIONAL FLOW, VORTICITY AND THE LIFE TIME OF SOLAR ACTIVE REGIONS}
\shortauthors{MAURYA and AMBASTHA}
\begin{document}

\title{SUB-SURFACE MERIDIONAL FLOW, VORTICITY AND THE LIFE TIME OF SOLAR ACTIVE REGIONS}

\author{R. A. MAURYA and A. AMBASTHA}
\affil{Udaipur Solar Observatory, Physical Research Laboratory, Udaipur-313001, India.}

\email{ramajor@prl.res.in, ambastha@prl.res.in}

\begin{abstract}
Solar sub-surface fluid topology provides an indirect approach to examine the internal characteristics of active regions (ARs). Earlier studies have revealed the prevalence of strong flows in the interior of ARs having complex magnetic fields. Using the Doppler data obtained by the Global Oscillation Network Group (GONG) project for a sample of 74 ARs, we have discovered the presence of steep gradients in meridional velocity at depths ranging from 1.5 to 5 Mm in flare productive ARs. The sample of these ARs is taken from the Carrington rotations 1980--2052 covering the period August 2001-January 2007. The gradients showed an interesting hemispheric trend of negative (positive) signs in the northern (southern) hemisphere, i.e., directed toward the equator. We have discovered three sheared layers in the depth range of \mbox{0--10 Mm}, providing an evidence of complex flow structures in several ARs. An important inference derived from our analysis is that the location of the deepest zero vertical vorticity is correlated with the remaining life time of ARs. This new finding may be employed as a tool for predicting the life expectancy of an AR.
\end{abstract}


%
\section{INTRODUCTION}
\label{sec:introduction}
Solar surface flows have been studied over the last three decades using mostly the surface observations such as the photospheric magnetic tracers and Doppler measurements. These tracers revealed that the magnetized regions rotate faster than the surrounding medium of field-free plasma \citep{1978ApJ...219L..55G,1993SoPh..147..207K,1996ARA&A..34...75H}.
Doppler measurements have shown poleward meridional flows at the solar surface \citep{1979SoPh...63....3D,1982SoPh...80..361L,1988SoPh..117..291U}. More recently, it has become possible to study sub-surface structures and flows subsequent to the advent of helioseismology \citep{1984ARA&A..22..593D,1991ARA&A..29..627G}, which probes the solar interior using acoustic modes of oscillations \citep{1970ApJ...162..993U,1971ApL.....7..191L}.
Helioseismic studies have revealed that sunspots are rather shallow, near-surface phenomena \citep{2000SoPh..192..159K,2004ApJ...610.1157B,2006ApJ...640..516C}, and their rotation rate with depth depends on the stage of evolution and age. \citet{2009A&A...506..875S} have reported that a majority of ARs displays sudden
decrease in the rotation speed, compatible with dynamic disconnection of sunspots from their parental magnetic roots.  Local helioseismology has further revealed the sunspots to be the locations of large flows near the surface \citep{2002ApJ...570..855H, 2004ESASP.559..337B, 2004ApJ...603..776Z}.

Earlier studies have found that magnetic helicity of ARs follows a hemispheric rule, i.e.,  positive (negative) in southern (northern) hemisphere \citep{2001ApJ...549L.261P}. Such a hemispheric trend is also reported for flows in ARs \citep{2005ApJ...631..636K, 2006SoPh..236..227Z, 2007ApJ...667..571K}. Zonal flows exhibit larger amplitudes in southern hemisphere especially at higher latitudes, coinciding with larger magnetic activity in that hemisphere. This may be attributed to the inclination of the solar rotation axis  \citep{2005ApJ...621L.153B}. Also, ARs advect poleward at nearly the same rate as the quiet regions (QRs) \citep{2009ApJ...698.1749H} and show convergent horizontal flows combined with cyclonic vorticity, counter clockwise in the northern hemisphere \citep{2004ApJ...605..554K, 2004ApJ...603..776Z}.

Activity related variations are also reported in solar surface flows  \citep{ 2002Sci...296..101V, 2003ApJ...585..553B, 2004ESASP.559..293A, 2005ApJ...634.1405H, 2009ApJ...706L.235M}. It is believed that sheared flows in sub-surface layers cause sunspot motions that may lead to unstable magnetic topologies, causing reconnection of magnetic field lines required for flares. \citet{2003ApJ...585..553B} have found that the maximum meridional velocity of surface flows is smaller when the Sun is more active. The maximum unsigned zonal and meridional vorticities of ARs are correlated with the total X-ray flare intensity \citep{2006ApJ...645.1543M}. Furthermore, steep meridional velocity gradients are found in flaring ARs at the depth range of \mbox{4--5 Mm} \citep{2004ESASP.559..293A, 2009ApJ...706L.235M},  which decreased after flares.

Although internal flows in ARs and QRs have been studied \citep{2006ApJ...645.1543M, 2009SoPh..258...13K, 2010mcia.conf..516M}, an understanding of their distinctive characteristics in flare productive ARs as compared to that in dormant ARs requires further investigation. In this letter, we report on the properties of internal flows in ARs of varying levels of flare productivity and magnetic complexity observed during the Solar Cycle 23, including large ARs such as NOAA 10030, 10484, 10486, 10070, etc., termed as super-active regions (SARs). These SARs are found to possess distinctly different characteristics of flows in their interiors as compared to the less productive or dormant ARs. We have addressed important issues on the relationship of flow characteristics with the life time of ARs, magnetic and flaring activities and  hemispheric trends.

%
%
%
\section{THE DATA AND ANALYSIS}
\label{S-DataAn}
For selecting our sample of the ARs, we have used archived information on ARs and solar activity provided by the web-pages of {\it Solar Monitor}\footnote{\url{http://www.solarmonitor.org}}. We first identified the ARs producing flares of X-ray class $>$ M1.0 using {\it GOES} database\footnote{\url{http://www.lmsal.com/SXT/plot_goes.html}} during Carrington rotations 1980--2052 of Solar Cycle 23. Then, we short-listed the ARs within the heliocentric location $\pm40$\arcdeg~to avoid projection effects and selected 74 ARs, both flaring and relatively dormant, that were well covered by the GONG network. For each flaring AR, we chose a single data set corresponding to the day of maximum flaring activity. (A full list of the selected ARs is not provided here because of the space limitation.)

We examined the sub-surface flows in the interior of the ARs using the ring data products provided by GONG \citep{2003ESASP.517..295H}. It utilizes 16\arcdeg$\times$16\arcdeg$\times$1664 min data-cubes, where the first two quantities correspond to the spatial dimensions in degrees and the third is the time duration  in minutes. The three-dimensional Fourier transform of the data-cube gives the power spectrum which exhibits a trumpet like structure in the [$k_x, k_y, \omega$] space. Slices of the trumpet at given frequencies $\omega$ render concentric rings in [$k_x, k_y$] plane corresponding to different $p$-modes \citep{1988ApJ...333..996H}. A wave propagating through the medium with horizontal flow is advected to increased (decreased) frequency depending on its propagation along (opposite) the flow. The frequency shift $\Delta\omega$ for acoustic waves, $\Delta\omega={\bf U}\cdot{\bf K}$, gives the distortion in the shape of the rings. Using a proper fitting technique, this distortion can be estimated and the corresponding flow velocity is determined. As the modes of different wavelengths are trapped at different depths, the flow patterns derived at the surface are the weighted average over depths. This concept is used for deriving the flow beneath the surface using regularized least square (RLS) inversion \citep{1985SoPh..100...65G} adopted for GONG data \citep{2003ESASP.517..255C}.

We can derive vertical vorticity, $\omega_z$,  to describe the circulation in ARs, as follows
\[
\omega_z=(\nabla\times u)_z=\frac{\partial u_y}{\partial x}-\frac{\partial u_x}{\partial y}.
\]
\noindent Here, $u_x$ and $u_y$ are the zonal and meridional components of flow, respectively, and $\omega_z$ is the vertical vorticity, which is a physical measure of vertical twist in the flow. In order to ascertain its significance as flow characteristic, we carried out an analysis of ARs with corresponding QRs located at the same latitude and time, but different longitudes. We found that the relative values of $u_y$ and $\omega_z$ in ARs were indeed significantly larger than in the QRs. It has been shown observationally \citep{2008A&A...477..285S, 2009A&A...506..875S} as well as theoretically \citep{2009ApJ...701.1300J} that Maxwell stresses play a significant role in modifying horizontal flows. \citet{2009NewA...14..429S} have also shown that flow parameters obtained for ARs are mainly influenced by their magnetic fields as compared to QRs where the flows correspond to supergranular structures. Therefore, it is clear that the flows below ARs are influenced ``magnetically''  that are otherwise driven purely by hydrodynamical processes, i.e., Reynolds stress and pressure gradient.

We examined the relationship of internal flow with magnetic activity of ARs using the magnetic index, MI, obtained from SOHO/MDI 96 minutes averaged magnetograms. For this,  we extracted the area of interest from the full disk magnetograms corresponding to the 16\arcdeg$\times$16\arcdeg~spatial patch used in the ring analysis. Then MI corresponding to the AR is calculated from the averaged absolute values of magnetic fields over the patch. We also calculated the flare index, FI, for each AR by multiplying the X-ray flux, obtained from GOES, with the flare duration and then summing the contributions from all the flares that occurred during the 1664 minutes' period of the data-cube.

We estimated the remaining life times ($\tau$) of ARs from the information provided by {\it solar monitor} for the Earthside and MDI farside imaging as follows. We followed the signature of an AR using magnetograms from the reference, central time $t_r$ of the data cube until the time $t_e$ when the AR disappeared. In case when the AR rotated past the west-limb of the Sun to the farside, we followed up its signature using the acoustic hologram images until its disappearance at time $t_e$. The remaining life time ($\tau$) of an AR is then estimated from the difference $(t_e-t_r)$.
%
%
%
\section{RESULTS AND DISCUSSIONS}
\label{S-ResCon}
\begin{figure}  
	\centering
		\includegraphics[width=0.95\textwidth,clip=,bb=23 20 435 297]{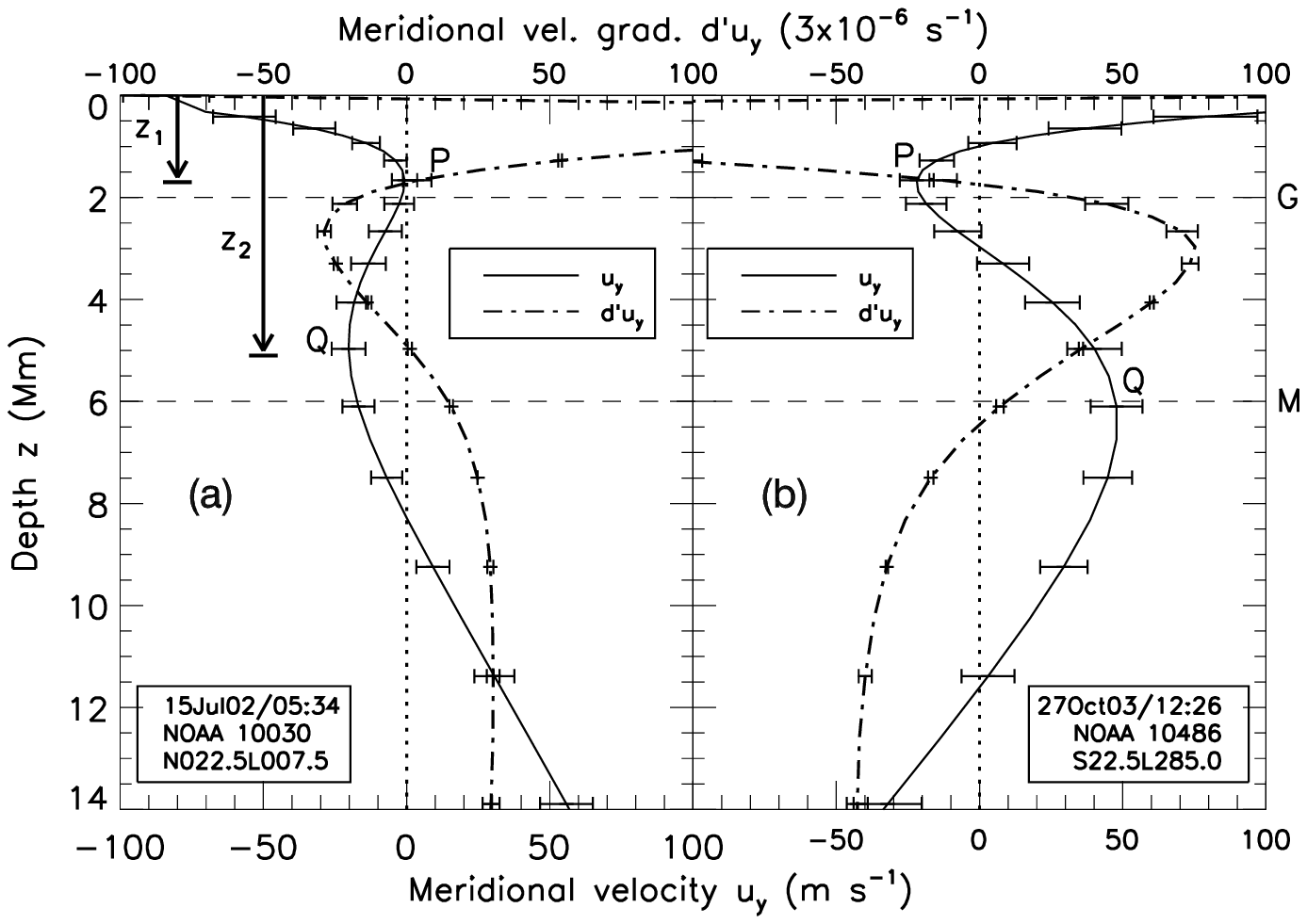}
		\caption{Meridional velocity profiles (solid curves) in the depth range 0--14 Mm for two ARs (a) NOAA 10030 (northern hemisphere) and (b) NOAA 10486 (southern hemisphere). The dash-dotted curves correspond to the meridional velocity gradients. Error bars show the corresponding uncertainties at 1$\sigma$ level. Horizontal dashed lines at the depths of 2 and 6 Mm, marked G and M, correspond to the convective scale sizes of granules and mesogranules, respectively.}
	\label{fig1}
\end{figure}

\begin{figure}  
	\centering
		\includegraphics[width=0.95\textwidth,clip=,bb=15 9 579 545]{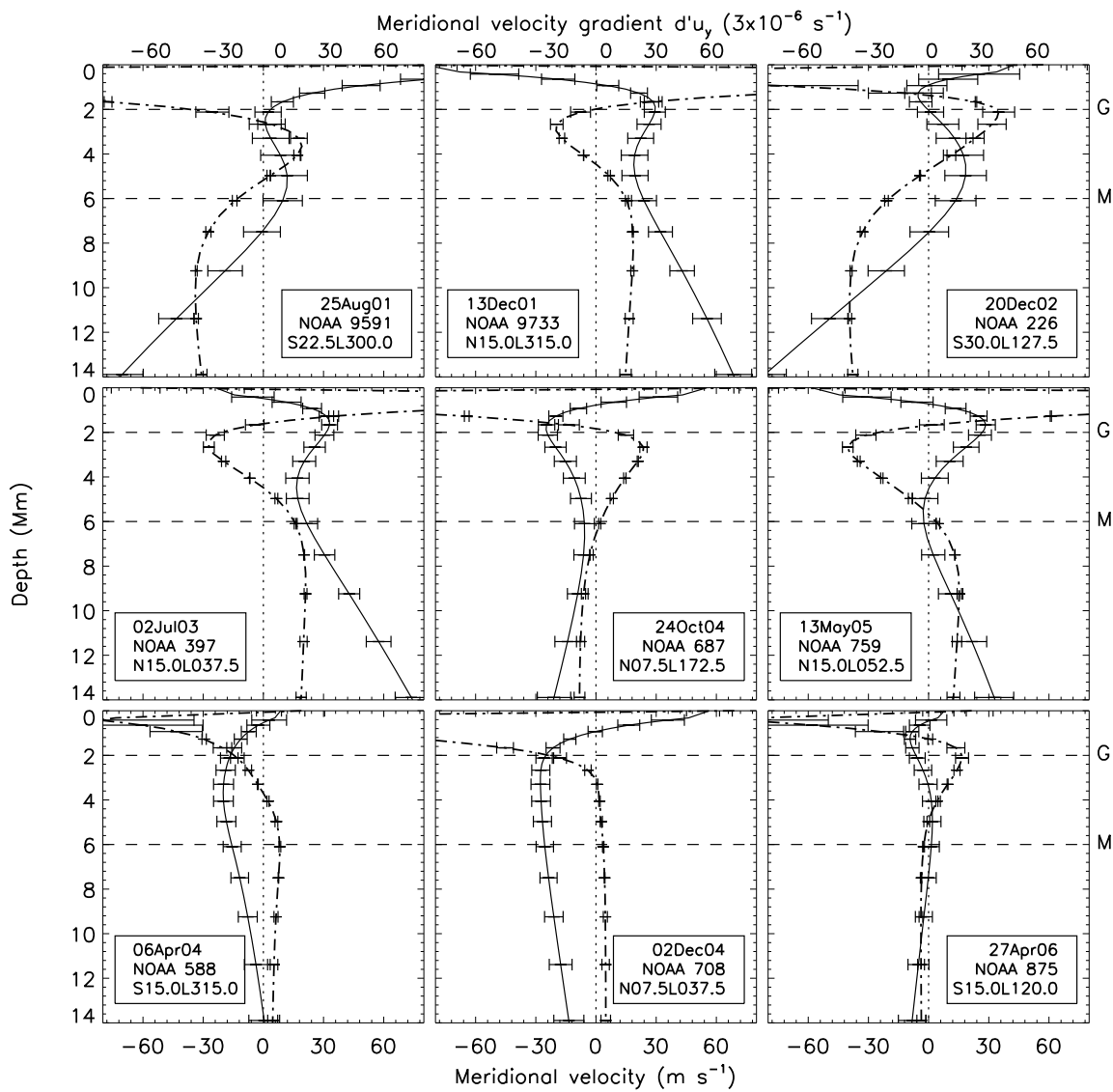}
		\caption{Meridional velocity profiles (solid curves) in the depth range 0--14 Mm for nine ARs of our sample. The dash-dotted curves correspond to the meridional velocity gradient. Error bars show the corresponding uncertainties at 1$\sigma$ level. Horizontal dashed lines at the depths of 2 and 6 Mm, marked G and M, correspond to the convective scale sizes of granules and mesogranules, respectively.}
	\label{fig2}
\end{figure}

\begin{figure}  
	\centering
		\includegraphics[width=0.72\textwidth,clip=,bb=18 1 350 565]{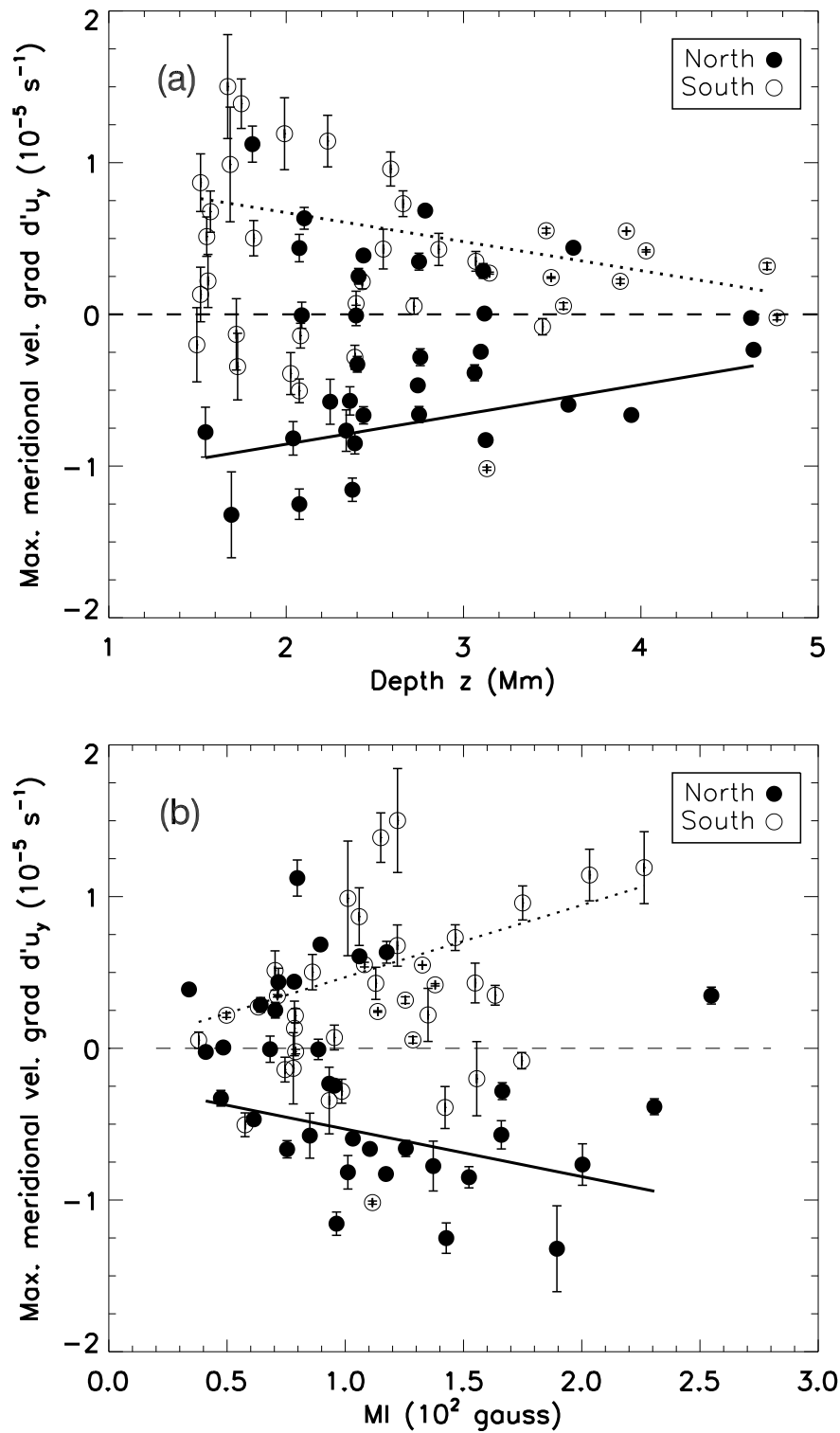}
		\caption{Distributions of the maximum meridional velocity gradient $d'u_y$ for 74 ARs with (a) depth z, and (b) magnetic index MI of the AR. Filled (open) circle represent the northern (southern) hemispherical ARs. The solid (dotted) fitted lines drawn through these points indicate the general hemispheric trend of the ARs.}
	\label{fig3}
\end{figure}

\begin{figure}  
	\centering
		\includegraphics[width=0.74\textwidth,clip=,bb=30 17 360 516]{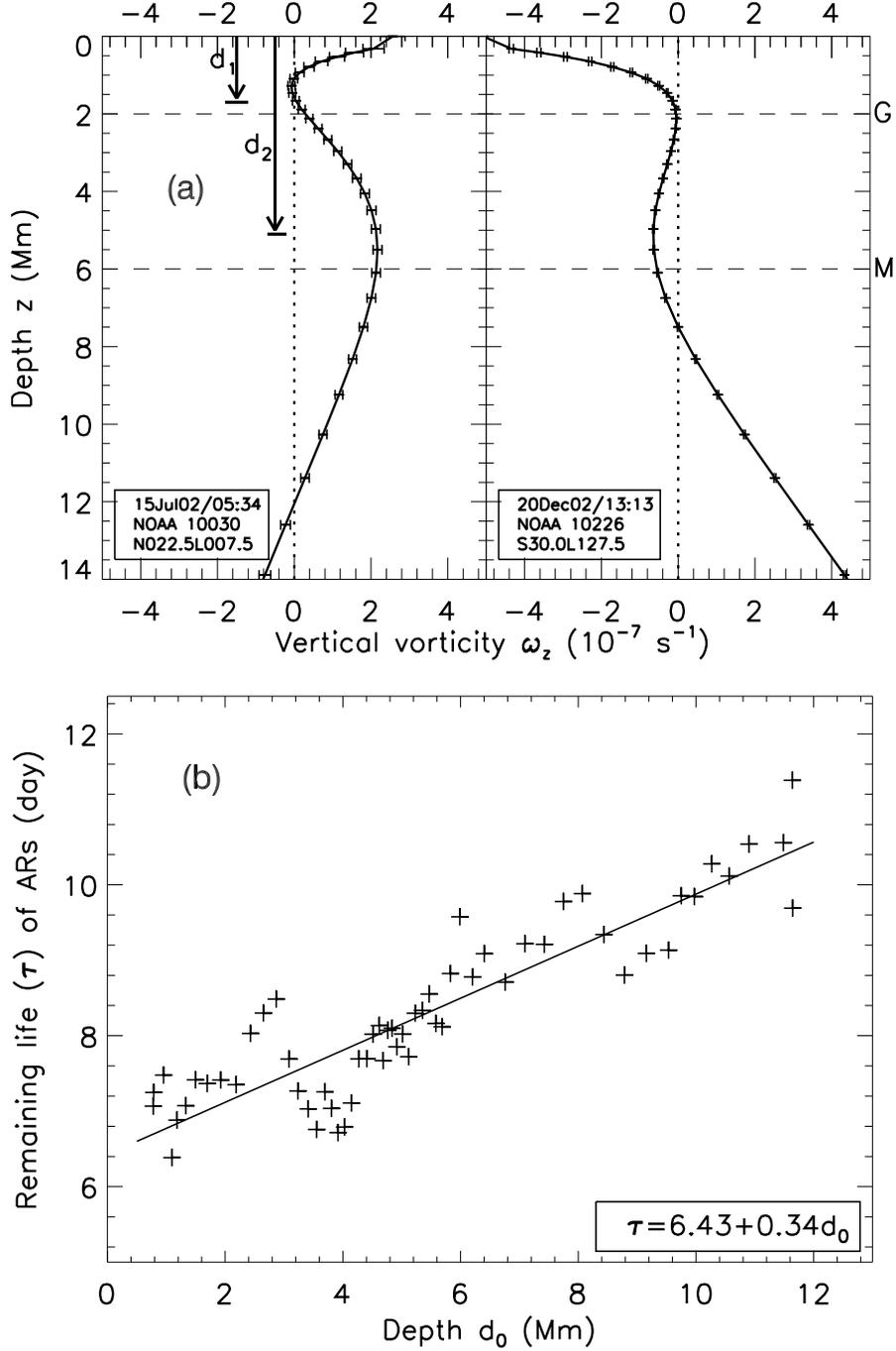}
		\caption{(a) Vertical vorticity profiles, with error bars, corresponding to two ARs, NOAA 10030 (northern hemisphere) and NOAA 10226 (southern hemisphere). Horizontal dashed lines at the depths of 2 and 6 Mm, marked G and M, correspond to the convective scale sizes of granules and mesogranules, respectively. (b) Variation of the remaining life time of the sample of 74 ARs with the depth of deepest zero vertical vorticity.}
	\label{fig4}
\end{figure}

The main results of our analysis for the sample of ARs are presented in Figures~\ref{fig1}--~\ref{fig4}. Meridional velocities, $u_y$, obtained for two of the ARs, NOAA 10030 (northern hemisphere) and 10486 (southern hemisphere), are shown by the solid curves in Figure~\ref{fig1}. The velocities are obtained from the surface to a depth of 14 Mm. The data sets used for these ARs correspond to 15 July 2002 and 27 October 2003, respectively. The dash-dotted curves show the greatest rate of change, or the gradient of $u_y$, along the vertical direction z, i.e., $du_y/dz \equiv d'u_y$. The magnitudes of $u_y$ and $d'u_y$ are evidently more significant as compared to the errors in the depth range from 1.5 to 10 Mm.

The profiles of $u_y$ and $d'u_y$ for the two ARs, located in opposite hemispheres, exhibit opposite trends with depth. The two extrema near P and Q of $u_y$, however, occurred at around the same depths for both the ARs. Evidently, $d'u_y$ changed sign with increasing depth from the surface to a depth of 1.5 Mm where $d'u_y$ is positive (negative) in the northern (southern) hemisphere. Thereafter, it changed sign again in the depth ranges of 1.5--5 Mm and 5--10 Mm. From these sign reversals of $u_y$ and $d'u_y$ profiles, we can infer that there exist three sheared layers in the interior of these two ARs. This appears to be a general feature of flare productive and complex ARs of our sample, some more of which are shown in Figure~\ref{fig2}.

We may infer that the depths of 2 and 6 Mm near which the extrema of $u_y$ are located correspond to the convective scale sizes of granulation and mesogranulation. For reference, these depth levels are marked in Figs.~\ref{fig1}--~\ref{fig2}  as ``G'' and ``M'', respectively. These depths were also referred as the locations of ionization zones of H$^+$ and He$^+$ ions by \citet{1964ApJ...140.1120S} and \citet{1981ApJ...245L.123N}. However, recently \citet{2001ApJ...563L..91C} and \citet{2003ApJ...597.1200R} have shown this argument to be incorrect. \citet{2000A&A...357.1063R} suggested that mesogranulations result from the non-linear interaction of granules \citep[also][]{2001ApJ...563L..91C, 2003ApJ...597.1200R}. \citet{2003ApJ...597.1200R} suggested that granular scales are determined by a close association between the upflows, which sustain radiative loss, and the downflow plumes, which initiate the upflow motions. A more detailed description of surface convection and scales can be found in the recent review article by \citet{2009LRSP....6....2N}.

Convective cells are parcels of fluid moving from the deeper layers to the upper layers and back to the bottom. The vertically rising material must change direction along a horizontal flow at two positions, i.e., the bottom and the top of convective cells. In a depth distribution of the velocity flow, these turning points may be represented by peaks in the velocity profile. The extrema seen in the vertical profiles of $u_y$, such as the one at ``G''  in Figure~\ref{fig1}, may be due to instabilities that drive the convective cells. The reason for convective driving near these depths is not clear because it depends on many factors, e.g., physical properties of flux rope and its ambient medium.

From our sample we found that 44 ARs displayed two extrema in their $u_y$ profiles while the rest, i.e., 30 ARs possessed a single extremum in the depth range 0--10 Mm. A statistical analysis of these ARs showed that more complex and flare productive ARs possessed two extrema, located at depths of 1.92$\pm$0.15 and 4.69$\pm$0.30 Mm, respectively. On the other hand, relatively dormant ARs possessed a single extremum located at a shallower depth of 1.66$\pm$0.97 Mm, i.e., with large depth uncertainty. There are some further interesting features associated with the ARs having two extrema as compared to the ARs having only one extremum: i) larger mean magnetic field, ii) nearly twice the mean GOES flux, and iii) nearly twice the mean life times.

The maximum $d'u_y$ of ARs in the two hemispheres exhibited a general trend, as evident from their distribution with depths (Figure~\ref{fig3}a).
Most ARs lying in the northern (southern) hemisphere possessed their maximum negative (positive) gradients in the depth range 1.5--6 Mm. There are 24 (70\%) out of the 34 ARs located in the northern hemisphere having negative maximum gradients while 29 (74\%) out of the 39 ARs located in the southern hemisphere having positive maximum gradients. We also noticed that some of those ARs which did not follow this hemispheric trend were new emerging ARs.

It is clear from Figure~\ref{fig3}(a) that the magnitude of maximum $d'u_y$ decreased with depth. The fits for ARs in northern (southern) hemisphere are drawn in solid (dotted) line. These lines are described by $d'u_y=-1.15+0.21z$ for the northern ARs and $d'u_y=1.05-0.19z$ for the southern ARs, where, $z$ represents the depth of the maximum $d'u_y$. We do not clearly understand the reason as to why the maximum $d'u_y$ should have such a linear relationship with depth, as only 25\% (21\%) of the total variance in $d'u_y$ in the northern (southern) hemisphere is explained by the linear regression model. However, the hemispheric trend is clearly evident from the averages of $d'u_y$ obtained for the northern and southern hemispheres, i.e., ($-5.87\pm0.72)\times10^{-4}\,{\rm s^{-1}}$ and ($5.86\pm0.70)\times10^{-4}\,{\rm s^{-1}}$, respectively.

The relationship of $d'u_y$ with magnetic index (MI) of ARs is shown in Figure~\ref{fig3}(b). Although the ARs show a rather large scatter, there still appears to be a correlation between these two parameters. The fitted lines drawn for the ARs in the northern and southern hemispheres are given by $d'u_y=-0.22-0.31$\,MI and $d'u_y=0.01+0.48$\,MI, respectively. These fits reveal that the maximum $d'u_y$ of ARs increased with MI. The larger slope of the fitted line corresponding to the ARs located in the southern hemisphere indicates larger $d'u_y$ for these as compared to the ARs of similar MI located in the northern hemisphere.

We have computed the vertical component of vorticity, $\omega_z$, for all the ARs of our sample, but show the profiles only for two ARs, NOAA 10030 and 10226 ( Figure~\ref{fig4}, top panel). It is evident that $\omega_z$ varies with depth both in sign and magnitude as the flow parameters shown in Figure~\ref{fig1}. These profiles also show two extrema of opposite signs at the depths near 2 and 6 Mm corresponding to the scales ``G'' and ``M''. This implies that the flows at these depths are twisted in opposite directions. For NOAA 10030, $\omega_z$ changed direction around the depths $\sim$1.8 and $\sim$12 Mm while for NOAA 10226, the corresponding depths are $\sim$2 and $\sim$8 Mm.

Zero vorticity may be a signature of the flux rope being broken at these depths. It is possible that the connection between the magnetic structures at the surface and its underlying roots may get broken through a dynamical disassociation process \citep{1994ApJ...436..907F}, and surface reconnection of the two opposite polarities \citep{1999SoPh..188..331S}. But these mechanisms have been overruled by dynamical disconnection mechanism proposed by \citet{2005A&A...441..337S} based upon the buoyant upflow of plasma along the field lines. From the simulation of thin flux tubes, \citet{2005A&A...441..337S} found that the disconnection takes place at a depth between 2 and 6 Mm, which correspond with the depths ``G'' and ``M'' marked in Figure~\ref{fig4}(a). We suggest that once the flux ropes in the interior of an AR are disconnected, its remaining life time would depend upon the depth where $\omega_z$ vanishes, i.e., $\omega_z$=0.

We have determined the depth of zero $\omega_z$ by a minimization procedure. However, it was not possible to accurately approach toward the point of zero vorticity for some of the ARs of our sample due to the limitation of the available spatial resolution of GONG data. For those ARs which did not show zero $\omega_z$ in their interior, we took its absolute minimum to relate with the remaining life time. Further, in Figure~\ref{fig4}(b), we have plotted the depths of zero vorticity, $d_0$, with the remaining life time, $\tau$. A good linear trend is evident between the parameters $d_0$ and $\tau$ from the fitted line, $\tau = 6.43 + 0.34\,d_0\,$, drawn through the boxcar averaged points. The fit is found to be good as 72.65\% of the total variance in $\tau$ is explained by the linear regression model.

Figure~\ref{fig4}(b) shows that ARs having zero vorticity at deeper levels last longer. The Pearson correlation coefficient between the depth and life time of ARs is $\sim$85\% for ARs in the northern and southern hemispheres. However, for the four SARs, i.e., NOAA 10030, 10044, 10069, 10486, 10488, the remaining life times are found to be $>$30 days, i.e., much larger than the rest of ARs of our sample. These SARs did not follow a linear relationship with depth unlike the others. This may be attributed to a continuous process of emergence of new flux tubes in these SARs resulting in enhancing their life. Due to their distinct abnormal behaviour, we have excluded these SARs from the fittings.

We have examined the relationship of flare productivity of ARs with the extrema of $\omega_z$. It is found that the depth of first extremum is mildly correlated with the integrated X-ray flux released during the 1664 minutes' period of the data cube obtained for the ARs. The Pearson correlation coefficient calculated between these parameters is 20$\%$. On the other hand, no correlation is found between the second extremum of $\omega_z$ and the integrated X-ray flux of the corresponding ARs.

\section{CONCLUSIONS}
\label{S-conc}
From our study of the 74 ARs observed during August 2001-January 2007 of Cycle 23, we have discovered statistically significant relationship among the sub-surface flow topology, energetics and the remaining life time of ARs. We have found the following important results:

\begin{enumerate}
	\item Three sheared layers in the depth range 0--10 Mm were found to exist in 44 ARs which revealed the complexity of flow structures beneath the surface of more complex ARs. The two extrema in $u_y$ and $d'u_y$ profiles of these ARs were found to be located at the depths of 1.92$\pm$0.15 and 4.69$\pm$0.30 Mm.
	\item The ARs having two extrema as compared to the ARs having only a single extremum in $u_y$ were found to be more active as they possessed as large as twice the mean magnetic field (MI), mean GOES X-ray flux and mean life time.
	\item The extrema of meridional velocity gradients were found to follow a hemispheric trend, viz., in the northern hemisphere, 24 (70\%) out of 34 ARs had negative gradients while in the southern hemisphere, 29 (74\%) out of 39 ARs had positive gradients.
	\item ARs of larger MI possessed steeper gradient in meridional velocity profiles.
	\item Flaring activity of an AR is found to be associated with depth of the first extremum of vertical vorticity.
	\item ARs having zero vertical vorticity at deeper layers are expected to last longer.
\end{enumerate}
	
In summary, we have discovered a new hemispheric trend involving the meridional velocity gradient of sub-surface flows. More importantly, life time of ARs appears to be correlated with the depth of the deepest zero vertical vorticity. This inference may be useful in predicting the expected life time of ARs.

%
\acknowledgments

This work utilizes data obtained by the Global Oscillation Network Group (GONG) program and the Solar Oscillations Investigation/Michelson Doppler Imager (SOI/MDI) on the Solar and Heliospheric Observatory (SOHO). Solar Monitor Data was used to find the correct location of the ARs. We would like to thank the referees for their comments and important suggestions that helped to improve the manuscript.



\newpage

\end{document}